\documentclass{amsart}
\usepackage{amsmath,amsfonts,amssymb,graphicx,setspace}
\usepackage{hyperref}

\def\bigS{ {\mathcal {S}}}

\def\real{\mathbb{R}}
\newcommand{\R}{\mathbb R}

\def\be#1\ee{\begin{equation}#1\end{equation}}
\newcommand{\fer}[1]{(\ref{#1})}

\newcommand{\bq}{\begin{equation}}
\newcommand{\eq}{\end{equation}}
\newenvironment{equations}{\equation\aligned}{\endaligned\endequation}
\def\bqa{\begin{eqnarray}}
\def\eqa{\end{eqnarray}}

\def\e{\epsilon}

\newcommand{\bd}{\begin{displaymath}}
\newcommand{\ed}{\end{displaymath}}
\newcommand{\ba}{\begin{eqnarray}}
\newcommand{\ea}{\end{eqnarray}}

\def\ff{\widehat f}

\def\gg{\hat g}

\def\R{\mathbb{R}}

\newcommand{\setR}{\mathbb{R}}

\theoremstyle{plain}

\title{Kinetic models for the trading of goods}

\author{Giuseppe Toscani, Carlo Brugna and Stefano Demichelis}

\thanks{Giuseppe Toscani is with the Department of Mathematics, Carlo Brugna is with the Department of Physics,
 Stefano Demichelis is with the
 Department of Economics and Business,
 of the University of Pavia, Italy.  e.mail: giuseppe.toscani@unipv.it, cabrugn@tin.it,
 stefano.demichelis@unipv.it.
}

\date{\today}

\begin{document}
\maketitle

\begin{center}\small
\parbox{0.85\textwidth}{

\textbf{Abstract.} In this paper we introduce kinetic equations for the
evolution of the probability distribution of two goods among a huge
population of agents. The leading idea is to describe the trading of these
goods by means of some fundamental rules in price theory, in particular by
using Cobb-Douglas utility functions for the binary exchange, and the
Edgeworth box for the description of the common exchange area in which
utility is increasing for both agents. This leads to a Boltzmann-type
equation in which the post-interaction variables depend in a nonlinear way
from the pre-interaction ones. Other models will be derived, by suitably
linearizing this Boltzmann equation. In presence of uncertainty in the
exchanges, it is shown that the solution to some of the linearized kinetic
equations develop Pareto tails, where the Pareto index depends on the
ratio between the gain and the variance of the uncertainty. In particular,
the result holds true for the solution of a drift-diffusion equation of
Fokker-Planck type, obtained from the linear Boltzmann equation as the
limit of quasi-invariant trades.
\medskip

 }
\end{center}

\section{Introduction}
In an effort to understand the emergent properties appearing in
complex agent-based systems, various concepts and techniques of
statistical mechanics have been fruitfully applied in the last
twenty years to both social and economic fields. This is well
documented both by recent books on these topics
\cite{Basu,book2,book1, NPT,book3,book4}, as well as by the
introductory articles \cite{review,DY00,Gup06, hayes, encyclopedia}.
In particular, the point of view of collisional kinetic theory
revealed to be a powerful instrument to describe the behavior of
systems in which the mechanism of variation could be described
mostly in terms of binary interactions between agents
\cite{CCM,CCS,IKR, CPT, slanina}. In kinetic theory of rarefied gas,
where binary collisions between molecules are the dominant
phenomenon, the evolution of the (spatially uniform) density of the
molecules is described by the spatially homogeneous Boltzmann
equation \cite{Cer, Cer94},
 \be\label{Bol}
 \frac{\partial f(v,t)}{\partial t} =
 Q (f,f)(v,t).
 \ee
In \fer{Bol} the bilinear $Q $ term accounts for all kinematically
possible (those that conserve both  momentum and energy) binary
collisions. The post--collisional velocities $(v^*,w^*)$ are
consequently given by the linear (in terms of the pre--collisional
ones $(v,w)$) relations
 \be\label{co1}
 v^* = {1\over 2}(v+w+|v-w|{\bf n}) ,\qquad w^* = {1\over 2}(v+w-|v-w|{\bf n}),
 \ee
in which ${\bf n}$ is a unit vector. The Boltzmann equation
\fer{Bol} describes the relaxation of the gas towards equilibrium,
which in this case is given by the Maxwellian density \cite{Cer,
Cer94} of mass $\rho$, velocity $U$ and temperature $T$, that is
 \begin{equation} \label{equi-M}
M(v) := \frac{\rho}{(2 \pi T)^{3/2}} \exp\left \{ - \frac{|v-
U|^2}{2 T} \right \}.
\end{equation}
One of the most used models for the collisional operator $Q$ is the
so-called Maxwellian molecules operator \cite{ BMP02, BK00, BK02,
BCT06, Bobylev-Carrillo-Gamba}, given by
 \be\label{max}
Q\bigl(f\bigr)(v) = \int_{{\R}^3\times{S^2}} B\left((v-w) \cdot {\bf
n}\right)\bigl[ f( v^*)g( w^*)] \, d w d{\bf n} - \rho f(v).
 \ee In \fer{max}
$B(\cdot)$ is a measure of the collision frequency, $d{\bf n}$ denotes the
{\it normalized} surface measure on the unit sphere $S^2$, and
$$\rho = \int_{\R^3}f(v)\, d v\ .$$
For Maxwellian molecules the operator simplifies, and the positive
part of it takes the form of a generalized convolution \cite{Bob}.

Boltzmann-like equations for the evolution in time of the density of
wealth in a market composed by many agents are easily recovered by
identifying \emph{molecules} with \emph{agents},
\emph{particles'energy} with \emph{agents' wealth}, and \emph{binary
collisions} with \emph{trade interactions}. In this way, the kinetic
description of market models via a Boltzmann-type equation provides
one possible explanation for the development of universal profiles
in wealth distributions of real economies.

To obtain the kinetic description one usually resorts to few reasonable
assumptions, that date back to the work of Angle \cite{angle}. First,
agents are considered indistinguishable.
 Then, an agent's \emph{state} at any instant of time $t\geq0$ is completely
characterized by his current wealth $w\geq0$. When two agents
encounter in a trade, their {\em pre-trade wealths\/} $(v,w)$ change
into the {\em post-trade wealths\/} $(v^*,w^*)$ according to the
(linear) rule
\begin{equation}
  \label{eq.trules}
  v^* = p_1 v + q_1 w, \quad w^* = q_2 v + p_2 w.
\end{equation}
In most of the existing models, the {\em interaction coefficients\/}
$p_i$ and $q_i$ are non-negative random variables. While $q_1$
denotes the fraction of the second agent's wealth transferred to the
first agent, the difference $p_1-q_2$ is the relative gain (or loss)
of wealth of the first agent due to market risks. One generally
assumes that $p_i$ and $q_i$ have fixed laws, which are independent
of $v$ and $w$, and of time.

In these one-dimensional models, the probability distribution $f(w,t)$ of
wealth of the ensemble coincides with agent density and satisfies the
associated spatially homogeneous Boltzmann equation of Maxwell type
\fer{max}
\begin{equation}
  \label{eq.boltzmann}
 \frac{\partial f}{\partial t}  = Q_+(f,f) -f ,
\end{equation}
on the real half line, $w\geq0$. The collisional gain operator
$Q_+(f,f)$, which quantifies the gain of wealth $v$ at time $t$ due
to binary trades, acts on test functions $\varphi(w)$ as
\begin{align}
\nonumber Q_+(f,f)[\varphi] :=&
  \int_{\setR_+} \varphi(w)Q_+\big(f,f\big)(w)\,dw\\
  \label{eq.qweak}
  =& \frac12 \int_{\setR_+^2} \langle \varphi(v^*) + \varphi(w^*) \rangle f(v)f(w)\,dv\,dw,
\end{align}
with $\langle\cdot\rangle$ denoting the expectation with respect to the
random coefficients $p_i$ and $q_i$ in (\ref{eq.trules}). The large-time
behavior of the density is heavily dependent of the evolution of the
average wealth \cite{DMTb,MTa}
\begin{equation}
  \label{eq.conserve}
  M(t) :=  \int_{\setR_+} w f(w,t)\,dw  ,
\end{equation}
Conservative models are such that the average wealth of the society
is conserved with time, $M(t) = M$, where the value of $M$ is
finite. In terms of the interaction coefficients, this is equivalent
to impose the conditions $\langle p_1+q_2\rangle = \langle
p_2+q_1\rangle = 1$ .

The Boltzmann equation \fer{eq.boltzmann} belongs to the Maxwell type. As
in operator \fer{max} the collision frequency is independent of the
relative velocity, and the loss term in the collision operator is linear.
This introduces a great simplification, that allows to use most of the
well established techniques developed for the three-dimensional spatially
homogeneous Boltzmann equation for Maxwell molecules in the field of
wealth redistribution \cite{CPT,DMT,DMTb,DT,MTa,MTb,PT}.

In general, the richness of the steady states for kinetic market
models is the main remarkable difference to the theory of Maxwell
molecules \cite{Bob}. While the Maxwell distribution \fer{equi-M} is
the universal steady profile for the velocity distribution of
molecular gases, the stationary profiles for wealth can be manifold,
and are in general not explicitly known analytically. In fact, they
depend heavily on the precise form of the microscopic modeling of
trade interactions. Consequently, in investigations of the
large-time behavior of the wealth distribution, one is typically
limited to describe a few analytically accessible properties (e.g.\
moments and smoothness) of the latter.

This richness of steady states makes both the theoretical and
numerical study of these models highly interesting. Indeed, it
corresponds mathematically to identify the limit distribution in a
problem which has many analogies with a generalized central limit
theorem. Also, it allows to clarify the types of binary interactions
which produce fat tails in the distribution at infinity (Pareto
tails), which are shown to form in the profiles of the distribution
of wealth in western societies.

On the other hand, the socio-economic behavior of a (real)
population of agents is extremely complex. Apart from elements from
mathematics and economics, a sound description --- if one at all
exists
--- would necessarily need contributions from various other fields,
including psychology. Clearly, the majority of the available
mathematical models are too simple to even pretend to reflect part
of the real situation.

For a better understanding of some of the main outcomes of markets
in western societies, in recent years the kinetic community started
to improve the economic modeling of agent systems by including more
realistic assumptions, without making too heavy the underlying
Boltzmann equations. Two papers in particular fit into this line of
thinking.

In \cite{CPP} Cordier,  Pareschi and Piatecki derive a kinetic
description of the behavior of a simple financial market where each
agent can create their own portfolio between two investment
alternatives: a stock and a bond. In this case, the variation of
density is not based on linear binary collisions like
\fer{eq.trules}, but it is derived starting from the more realistic
Levy-Levy-Solomon microscopic model for price formation
\cite{LLS,LLSb}. The model in \cite{CPP} consists of a linear
Boltzmann equation for the wealth distribution of the agents coupled
with an equation for the price of the stock. From this model, under
a suitable scaling,  Cordier,  Pareschi and Piatecki derive a
Fokker-Planck equation and show that the equation admits a
self-similar lognormal behavior. For the first time, the kinetic
model in \cite{CPP} attempts to join to simple financial rules a
kinetic equation of Boltzmann type, able to describe a complex
behavior that could then mimic the market and explain the price
formation mechanism.

A second interesting example of this coupling has been recently
proposed in \cite{MD} by Maldarella and Pareschi. They introduce a
relatively simple model for a financial market characterized by a
single stock or good and an interplay between two different trader
populations, chartists and fundamentalists, which determine the
price dynamics of the stock. The model has been inspired by the
microscopic Lux--Marchesi model \cite{LMa,LMb}. The main novelty
here is to couple the financial rules with the opinion of traders
through a kinetic model of opinion formation recently introduced in
\cite{To1}. Moreover, some psychological and behavioral components
of the agents, like the way they interact with each other and
perceive the risk, which may produce nonrational behaviors, are
taken into account. This is done by means of a suitable \emph{value
function} in agreement with the Prospect Theory by Kahneman and
Tversky \cite{KTa,KTb}. As they show, people systematically
overreacting produces substantial instabilities in the stock market.

These two papers indicate that the methods of kinetic theory can be
fruitfully used to understand the market behavior even in presence
of more realistic ways of interactions between agents. The objective
weakness of models based on linear binary trades of type
\fer{eq.trules} is linked to the fact that people trades according
to rules which are in general difficult to justify from a
microeconomic point of view. The common ingredients which appear in
these trades are the saving propensity concept \cite{CC00} (agents
usually do not trade with all their wealth), and the risk concept
\cite{CPT} (there is no certainty of gain). While highly reasonable,
this type of interaction does not clarify while agents are pushed to
trade.

One of the fundamental assumption of prize theory in economics is
that people trades to improve its utility. Among binary interactions
which are motivated by an increasing of the individual utility, one
of the most used pictures is furnished by the Edgeworth box
\cite{Edg}, which is frequently used in general equilibrium theory,
and can aid in representing the competitive equilibrium of a simple
system or a range of such outcomes that satisfy economic efficiency.
Edgeworth box can fruitfully be applied in presence of an
agent-based system in which agents possess a finite number of goods
of $n\ge 2$ different types. Inspired by this mechanism of
increasing utility and competitive equilibrium, we will introduce in
the following a kinetic equation of Boltzmann type for the evolution
of the distribution density of the quantities of two goods, said
$(x, y)$, in a system of agents. In other words, we will be
concerned with the time evolution of the probability density
$f(x,y,t)$ of having $x$ goods of the first type and $y$ goods of
the second type at time $t >0$, given a known initial distribution
of these goods at time $t=0$. The variation in time of $f(x,y,t)$
will be entirely due to the microscopic binary interaction driven by
the Edgeworth box structure. We will take into account the
incomplete knowledge of the game in the binary interaction by
allowing agents to be wrong with respect to the ideal outcome
predicted by the increase of utility. This will be done by
introducing a randomness in the Edgeworth box outcome.

In more details, in the next Section we will resume the Edgeworth box
interaction, and the consequent Boltzmann-type equation, together with its
main properties. Surprisingly, on the contrary to most of the previous
types of trade, this exchange rule leads to a highly nonlinear binary
interaction, which is difficult to handle, if not numerically. For this
reason, in Section \ref{sec-lin} we will resort to a suitable linear
Boltzmann equation, which is obtained by allowing the agent to interact
(according to Edgeworth box), simultaneously with a sufficiently high
number of agents in the market. By Fourier based methods, it will be shown
that, under reasonable conditions on the initial data, this linear
equation has a unique solution, and the steady states are concentrated
along a well-defined line. Finally, in Section \ref{sec-FP}, we will
resort to an asymptotic procedure (the so-called \emph{quasi invariant
trade limit}), to obtain a linear Fokker-Planck equation which describes
the essentials of the trading of goods, and it is relatively easy to treat
from a mathematical point of view. In particular, it will be shown that
the solution converges towards a steady state with fat tails.

\section{The Edgeworth box}\label{Edge}
As discussed in the Introduction, most of the existing kinetic models for
wealth distribution are based on rigid assumptions which, if on one hand
can be shared, from the other hand are not deeply related to economic
principles, like price theory \cite{Fri}. The aim of this Section is to
introduce a new framework for trades, which is derived directly from the
basic principles of economy.

Individuals exchange goods. The benefits they receive depend on how
much they exchange and on what terms. Price theory tries to give an
answer to this fundamental question. In the case of a binary trade,
there may be many different exchanges, each of which would be
beneficial to both parties; some exchanges will be preferred by one
person, some by the other. There are then two different questions to
be settled. One is how to squeeze as much total gain as possible out
of the opportunities for trade; the other is how that gain is to be
divided. The two individuals who are trading have a common interest
in getting as much total gain as possible but are likely to disagree
about the division. Let us suppose in the following that agents in
the system possess the same two types of goods, and there is no
production, so that the total amount of goods remains unchanged. If
$(x_A, y_A)$ ($(x_B, y_B)$ respectively) denote the quantity of
goods of two agents $A$ and $B$, the quantities
 \be\label{per}
p_A = \frac{x_A}{x_A + x_B}, \quad q_A = \frac{y_A}{y_A + y_B}
 \ee
are the percentages of goods of the first agent. Note that the point
$(p_A,q_A)$ belongs to the square $\mathcal{S}= [0,1]\times[0,1]$. In
order to give a meaning to the reasons of trading,  it is classical to
assume that agent's behavior is driven by a utility function.  One of the
most popular of these functions is the {Cobb-Douglas utility function}
 \be\label{CD}
 U(p,q) = p^\alpha q^\beta, \quad \alpha + \beta = 1.
 \ee
Each agent will tend to {maximize its utility} by trading. The values
$\alpha$ and $\beta$ are linked to the preferences that the agent assigns
to the two goods. If $\alpha >\beta$, the agent prefers to possess goods
of the first type (numbered by $x$). The choice $\alpha= \beta = 1/2$
clearly means that the two goods are equally important for $A$. Given the
percentage point $(p_A,q_A)$ of the agent $A$, the curve
 \[
U(p,q) = U(p_A,q_A)
 \]
denotes the \emph{indifference curve} for the agent $A$. Indeed, any point
on the indifference curve has for $A$ the same utility. Note that the
indifference curve for $A$ entirely belongs to $\bigS$, and splits the
square into the two regions
 \[
U^-_A = \left\{(p,q) :U(p,q) < U(p_A,q_A)\right\}, \quad U^+_A =
\left\{(p,q) :U(p,q) > U(p_A,q_A)\right\}.
 \]
Clearly, any trade which will move the percentages of agent $A$ into the
region $U^+_A$ will increase its utility function, and will be acceptable
for the agent.

In the case of a binary trade, also agent $B$ has a Cobb-Douglas utility
function, with parameters of preference which in general are different
from the parameters of agent $A$. Likewise, it will be an indifference
curve for $B$, and a good region in which the utility of $B$ is increased
after trading. Ultimately, a trade will be acceptable for both agents, if
their percentages after the trade belong to the regions in which their
utilities are increased.

An ingenious way of looking at such a situation is the Edgeworth Box,
named after Francis Y. Edgeworth, the author of a nineteenth century work
on economics called Mathematical Psychics \cite{Edg}.

The Edgeworth box is built up by rotating the square $\bigS$ in which is
designed the indifference curve of agent $B$ of $180^\circ$ around the
center of the square $(1/2,1/2)$, and considering together the
indifference curve of $A$ and the rotated indifference curve of $B$ on the
same square.

Any point inside the box represents a possible division of the total
quantity of $A$ and $B$, with the share of $A$ measured from the lower
left-hand corner, and the share of $B$ from the upper right-hand corner.
Any possible trade is represented by a movement from one point in the box,
to another.

By construction, since the Cobb-Douglas utility function is jointly
convex, the utility of $A$ increases if the point moves up and to the
right; the utility of $B$ increases if the point moves down and to the
left. The region in which this happens is exactly the region common to
$U^+_A$ and to $U^+_B$ rotated of $180^\circ$.

This operative way of trading furnished by the Edgeworth box, has
interesting consequences. If the percentages of $A$ and $B$ are such
that there is a possible region for trading, once the trade has been
done, the point has moved inside the region, and the new region for
trading is smaller. If the two agents were so smart to choose a
point inside the region in which the two indifference curves are
tangent each other, since they curve in opposite directions, this
means that starting from this point, any point that is on a higher
indifference curve for the first agent must be on a lower curve for
the second; any trade that makes one better off makes the other
worse off. The point in which the two curves are tangent each other
is not unique. The set of all points from which no further mutually
beneficial trading is possible (Pareto optimal points) is called the
\emph{contract curve}.

The contract curve is related to prices. At that tangency of the two
indifference curves, in fact,  the slope of the tangency line
represents the relative prices for the two goods. Hence, there are
relative prices that will be consistent with the Pareto optimum, and
these prices will maximize the possible budget for both agents.

This short discussion on utility functions, trading and prices is
the basis of our construction of the binary trade. In what follows,
we suppose that all agents in the system have the same utility
function \fer{CD}. If agent $A$ has percentages $p$ and $q$ in his
Hedgeworth box, we consider as possible trades the movements of the
point $(p,q)$ into $(p^*,q^*)\in \bigS$, where
\begin{equations}\label{tr}
&p^* = p + {\lambda \beta (q-p) + \mu (q-p)} \\
    &q^* = q + {\lambda \alpha (p-q) + \tilde\mu (p-q)}.
\end{equations}
In \fer{tr}  $0 <\lambda \le 1$, and $\mu$ and $\tilde \mu$ are random
variables with zero mean and {finite variance}. We will assume moreover
that
 \be\label{co3}
0 <\lambda \beta  + \mu < 1, \quad 0 < \lambda \alpha  + \tilde\mu
\le 1.
 \ee
Under this condition, the post-trade point $(p^*, q^*)$ belongs to
$\bigS$, and it is an admissible point for the Edgeworth box. It is
immediate to show that, unless $p=q$, and in absence of randomness, that
is if
\begin{equations}\label{tr1}
&p^* = p + \lambda \beta (q-p)  \\
    &q^* = q + \lambda \alpha (p-q),
\end{equations}
the trade {increases the Cobb-Douglas utility function} of agent $A$.
Indeed
\[
\frac d{d\lambda} U( p^*, q^* ) = \alpha\beta (1-\lambda) (p-q)^2
p^{\alpha -1}  q^{\beta -1}{> 0},
\]
which, coupled with the convexity of the utility function gives
 \[
U( p^*, q^* ) = U(p,q) + \alpha\beta (p-q)^2 p^{\alpha -1} q^{\beta -1}
\lambda + \frac 12 \frac{d^2}{d\lambda^2} U(\tilde p^*, \tilde q^*
)\lambda^2 \ge
 \]
  \[
 U(p,q) + \alpha\beta (p-q)^2 p^{\alpha -1} q^{\beta -1} \lambda .
 \]
Note moreover that the trade \fer{tr1} is such that $p^* \not= q^*$,
unless $p=q$. Concerning agent $B$, since it starts from the point
$(1-p,1-q) \in \bigS$, according to \fer{tr1} it moves to the point
 \begin{equations}\label{tr2}
&\bar p = 1- p + \lambda \beta (1-q- (1- p))  \\
    &\bar q = 1- q + \lambda \alpha (1- p-(1- q)),
\end{equations}
which, proceeding as before, implies
 \[
U( \bar p, \bar q ) > U(1-p, 1-q).
 \]
Hence, in absence of randomness, the trade \fer{tr1} is such that
the Cobb-Douglas utility functions of both agents are increased,
locating the new point in $\bigS$ in the interior of the allowed
area for trades predicted by the Edgeworth box. Clearly, this trade
refers to the purely theoretical situation in which the traders $A$
and $B$ know all about the trade and its result. In general, this is
not realistic, and what traders can hope, after a certain number of
trades,  is that the majority of their trades had the result of
increasing their utility. This can be expressed by saying that a
reasonable result for traders is to obtain, in a single trade, the
increasing of their utility only in the mean. If one agrees with
this, then trades of type \fer{tr} satisfy all constraints of
Edgeworth box, but the single trade can move the point in a region
which is convenient only for one of the two traders, as well as in
the region which is not convenient for both.

Once the trade has been done, according to the rules of price theory, the
quantities of goods of agents $A$ and $B$ have been changed accordingly.
Making use of the constraints $x_A + x_B= x^*_A + x^*_B$ and $y_A + y_B=
y^*_A + y^*_B$, \fer{tr} implies for the agents $A$ and $B$ the post-trade
quantities of goods
 \begin{equations}\label{trA}
  &x^*_A = x_A + ({\lambda \beta + \mu}) \left( \frac{x_A +x_B}{y_A + y_B} y_A - x_A\right) \\
   & y^*_A = y_A + ({\lambda \alpha + \tilde\mu}) \left( \frac{y_A +y_B}{x_A + x_B} x_A - y_A\right)
   \\
  &x^*_B = x_B + ({\lambda \beta + \mu}) \left( \frac{x_A +x_B}{y_A + y_B} y_B - x_B\right) \\
    & y^*_B = y_B + ({\lambda \alpha + \tilde\mu}) \left( \frac{y_A +y_B}{x_A + x_B} x_B - y_B\right),
 \end{equations}
We remark that, on the contrary to what is usually assumed in the
wealth trades of type \fer{eq.trules}, the post-trade quantities in
\fer{trA} are related to the pre-trade quantities by nonlinear
relations. It can be verified, however, that the constraints about
the conservation of the total quantities of goods in the trade are
verified, and
 \[
x^*_A + x^*_B = x_A + x_B, \quad y^*_A + y^*_B = y_A + y_B .
 \]

\section{A linear Boltzmann equation for trading of goods}\label{sec-lin}
Let  $f(x,y,t)$ denote the {density of agents with quantities $x$ and $y$
of the two goods at time $t \geq 0$}. Without loss of generality, we will
assume in the following that $(x,y)$ are nonnegative real numbers.  A
Boltzmann-like equation of Maxwell type for the time evolution of
$f(x,y,t)$ can be written by observing that the \emph{collision rules} are
here given by \fer{trA}. A useful way of writing the Boltzmann equation
has been recalled in the introduction, where the Boltzmann equation
\fer{eq.boltzmann} has been rewritten resorting to the weak form
\fer{eq.qweak}. It corresponds to write, for any given smooth function
$\varphi$, the spatially homogeneous Boltzmann equation in the form
 \be\label{we}
\frac{d}{dt} \int_{\real_+^2}  \varphi(x, y) f(x,y,t)\,dx \, dy =
\int_{\real_+^2}\varphi(x ,y)Q(f,f)(x,y) \,dx dy .
 \ee
 The right-hand side of equation \fer{we} describes the change of
 density due to collision of type \fer{trA}. Given two agents with
 quantities of goods given by $(x, y)$ and $(x_1, y_1)$,
 one obtains from \fer{trA} the post-trade quantities
\begin{equations}\label{trB}
  &x^* = x + ({\lambda \beta + \mu}) \left( \frac{x +x_1}{y + y_1} y - x\right) \\
   & y^* = y + ({\lambda \alpha + \tilde\mu}) \left( \frac{y +y_1}{x + x_1} x -
   y\right).
   \end{equations}
Hence
\begin{equations}\label{qw}
& \int_{\real_+^2}\varphi(x ,y)Q(f,f)(x,y) \,dx  dy =
\\
&\left\langle \int_{\real_+^4}  ( \varphi(x^* ,y^*)-\varphi(x
,y))f(x,y,t)f(x_1,y_1,t) \,dx \, dy \, dx_1 \, dy_1 \,\right\rangle
.
 \end{equations}
In alternative, one can use the identity
\begin{equations}\label{qw1}
&\int_{\real_+^2}\varphi(x ,y)Q(f,f)  \,dx  dy  = {\frac 12}
\left\langle \int_{\real_+^4}  ( {\varphi(x^* ,y^* +
\varphi(x_1^* ,y_1^*)}\right. \\
 &{-\varphi(x ,y) - \varphi(x_1
,y_1)})f(x,y,t)f(x_1,y_1,t) \,dx \, dy \, dx_1 \, dy_1 \,\Big\rangle
.
 \end{equations}
By choosing $\varphi(x, y) = x$ (respectively $\varphi(x, y) = y$)
one verifies that the mean values
\begin{equations}\label{mv}
   & m_x(t) =  \int_{\real_+^2}  x f(x,y,t)\,dx \, dy \\
    &  m_y(t) =  \int_{\real_+^2}  y f(x,y,t)\,dx \, dy ,
\end{equations}
remain {constant in time} $m_x(t) = m_x(0)$ (respectively $m_y(t) =
m_y(0)$ ). These conservations are consequence of the constraints of
conservation of the quantities of goods in the trades. Since the exchanges
of goods of type \fer{trB} are nonlinear, the study of the properties of
the solution to the Boltzmann equation \fer{we} is difficult. In
particular, it is cumbersome to find the evolution of the higher moments
of the solution in a closed form. Hence, the Boltmann equation \fer{we} is
the starting point for a numerical study of the evolution of the density
by means of MonteCarlo methods \cite{BT,Pa}.

In order to derive a more treatable model,  which maintains most of the
features of the nonlinear one described by equation \fer{we}, let us study
in more details the exchange rule driven by the Edgeworth box, assuming
that, instead of single binary trades,  any agent is allowed to exchange
goods with part of the agent market, composed by a certain number of
agents. We remark that the exchange rule \fer{trA} for the quantity of the
first good can be fruitfully rewritten as
\begin{equation}\label{me1}
  x^*_A = x_A + ({\lambda \beta + \mu}) \left( {\frac{x_A +x_B}{y_A + y_B}} y_A - x_A\right) =
   x_A + ({\lambda \beta + \mu}) \left( {\frac{\frac{x_A +x_B}2}{\frac{y_A + y_B}2}} y_A -
   x_A\right).
\end{equation}
This writing clarifies that, in the post-trade quantity of goods in
\fer{me1}, the coefficient of $y_A$ is the {ratio between the mean
values} of the quantities $x$ and $y$ of the two agents exchanging
goods.

If one agent exchanges goods, according to the Edgeworth box
strategy,  with a number of $N$ agents, the quantity of the first
good changes according to
 \[
x^*_A = x_A + ({\lambda \beta + \mu}) \left( {\frac{x_A +x_1+ \cdots
+x_N}{y_A + y_1 +\cdots + y_N}} y_A -
  x_A\right),
 \].
and, if  $N$ is sufficiently large,
 \be\label{lin1}
 \frac{x_A +x_1+
\cdots +x_N}{y_A + y_1 +\cdots + y_N} = \frac{\frac{x_A +x_1+ \cdots
+x_N}{N+1}}{\frac{y_A + y_1 +\cdots + y_N}{N+1}} { \cong \frac
{m_x}{m_y}}.
  \ee
Since the {ratio between the mean values} is a conserved quantity,
by using approximation \fer{lin1} we obtain a consistent linear
version of the trade of goods, clearly valid under the assumptions
specified above. Therefore, by allowing agents to interact with part
of the market, according to the rules of the Edgeworth box, the
result of the trade can be well approximated by the linear exchange
rules
 \begin{equations}\label{lin2}
    &x^* = x + ({\lambda \beta + \mu}) \left( {\frac {m_x}{m_y}} y - x\right) \\
   & y^* = y + ({\lambda \alpha + \tilde\mu})\left( {\frac {m_y}{m_x}} x -
   y\right),
\end{equations}
in which the action of the market appears only through the mean values
$m_x$ and $m_y$. Consequently, if $f(x,y,t)$ denotes the {density of
agents with quantities $x$ and $y$ of the two goods at time $t \geq 0$},
$f(x,y,t)$ satisfies now the linear Boltzmann equation
 \be\label{Bl}
\frac{d}{dt} \int_{\real_+^2}  \varphi(x, y) f(t)dx dy =
\left\langle \int_{\real_+^2}  ( \varphi(x^* ,y^*)-\varphi(x
,y))f(t)dx dy \right\rangle .
 \ee
Also for equation \fer{Bl}, it can be easily verified that the mean
values \fer{mv} remain {constant in time}.

Due to its linearity, the Boltzmann equation \fer{Bl} can be studied
by resorting to Fourier transform, by standard methods available for
the Boltzmann equation for Maxwell molecules, and their extensions
to kinetic models of wealth distribution \cite{DMTb,MTa}. Among
other properties, the analysis of \cite{MTa} in terms of
Fourier-based metrics, allows to prove existence (and uniqueness) of
a steady distribution $f_\infty(x,y)$, under few reasonable
conditions on the initial distribution of goods, like existence of a
certain number of principal moments.

A quite useful expression for the Fourier transformed equation
follows by  writing the {linear} interaction \fer{lin2} as
\begin{equations}\label{new1}
  &m_y x^* = m_yx + ({\lambda \beta + \mu}) \left({m_x}y - m_yx\right) \\
   & m_xy^* = m_xy + ({\lambda \alpha + \tilde\mu})\left( {m_y} x -
   m_xy\right).
\end{equations}
The trade \fer{new1} corresponds to a dissipative interaction of
particles with a random addition proportional to the relative
velocity, similar to the one studied in \cite{CCT}. Then, the pair
$(x' =m_yx, y'= m_xy)$ can be fruitfully used as new set of
variables, and $f(x',y',t)$ still satisfies the Boltzmann equation
\fer{Bl}. By choosing
 \[
\varphi(x,y) = \exp\left\{ -i(x\xi + y \eta)\right\},
 \]
and using \fer{new1}, one directly obtain from \fer{Bl} the equation for
the Fourier transform $\ff$ of $f$, that takes the form
 \be\label{Bf}
\frac{\partial\ff(\xi,\eta)}{\partial t} = \langle \ff(\xi^*,
\eta^*)  \rangle - \ff(\xi,\eta).
 \ee
In \fer{Bf}, the post-interaction variables are given by
\begin{equations}\label{new4}
  &\xi^* =(1- ({\lambda \beta + \mu}))\xi + ({\lambda \alpha + \tilde\mu})\eta \\
   & \eta^* = ({\lambda \beta + \mu})\xi + (1-({\lambda \alpha + \tilde\mu}))\eta.
\end{equations}
Note that, by conditions \fer{co3}, the random coefficients in
\fer{new4} are non negative. It is immediate to check that the
post-trade Fourier variables satisfy the constraint
 \be\label{co2}
 \xi^* + \eta^* = \xi + \eta.
 \ee
This implies that the eventual stationary solutions are of the form
 \be\label{sta1}
 \ff_\infty(\xi, \eta) = \ff_\infty(\xi + \eta).
 \ee
Hence, the solution to the Boltzmann equation \fer{Bl} will
concentrate for large times along the line ${m_x}y = m_yx$.

To verify that this behavior holds true, one resorts to previous
works on similar subjects. The standard Fourier-based norm to be
used in connection with this type of equations \cite{DMTb} is
 \be\label{norm}
d_s(f,g) = \sup_{\xi,\eta \not= 0} \frac{|\ff(\xi,\eta) - \gg(\xi,
\eta)|}{(|\xi|^2 + |\eta|^2)^{s/2}}, \quad s>0.
 \ee
Relationships of this norm with other equivalent norms widely used in
probability theory and mass transportation can be found in \cite{CT}.
Uniqueness follows by the following argument. Let $\ff_1(\xi,\eta, t)$ and
$\ff_1(\xi,\eta, t)$ two solutions to equation \fer{Bf} corresponding to
the initial data $\ff_{0,1}(\xi,\eta)$ and $\ff_{0,2}(\xi,\eta)$. Then it
holds
 \[
 \frac{\partial }{\partial t}\frac{\ff_1(\xi,\eta, t) - \ff_2(\xi,\eta,t)}{(|\xi|^2 + |\eta|^2)^{s/2}}
  +  \frac{\ff_1(\xi,\eta, t) - \ff_2(\xi,\eta, t)}{(|\xi|^2 + |\eta|^2)^{s/2}}
 =
 \]
 \[
\left\langle \frac{\ff_1(\xi^*,\eta^*, t) - \ff_2(\xi^*,\eta^*,
t)}{(|\xi^*|^2 + |\eta^*|^2)^{s/2}} \frac{(|\xi^*|^2 +
|\eta^*|^2)^{s/2}}{(|\xi|^2 + |\eta|^2)^{s/2}}\right\rangle \le
 \]
 \[
d_s(f_1,f_2)\sup_{\xi,\eta \not= 0}\left\langle \frac{(|\xi^*|^2 +
|\eta^*|^2)^{s/2}}{(|\xi|^2 + |\eta|^2)^{s/2}}\right\rangle
 \]
Suppose that
 \be\label{c1}
C_s = \sup_{\xi,\eta \not= 0}\left\langle \frac{(|\xi^*|^2 +
|\eta^*|^2)^{s/2}}{(|\xi|^2 + |\eta|^2)^{s/2}}\right\rangle <+\infty
 \ee
Then, application of Gronwall's inequality gives (cfr. \cite{GTW95})
 \be\label{decay}
 d_s(f_1(t), f_2(t)) \le d_s(f_{0,1}, f_{0,2})\exp \left\{\left( C_s - 1)\right)t \right\}.
 \ee
From \fer{decay}, uniqueness of solution in a fixed time interval
follows. The case in which $C_s <1$ leads to better decay. In fact,
if this is the case, for all times $t \geq 0$ the $d_s$-distance of
$f_1$ and $f_2$ decays exponentially in time, and, proceeding as in
\cite{MTa} it can be proven that, if $f(x,y,t)$ is a weak solution
of the Boltzmann equation \fer{Bl}, which has initially finite
moments up to order $2$, then $f$ converges exponentially fast in
$d_s$ to a steady state $f_\infty$. In addition $f_\infty$ has the
same mean of the initial datum, and it is the only steady state with
this mean wealth.

To verify that the constant $C_s$ can be strictly less than one, it is in
general not direct. However, the following argument shows that we can
easily prove that the $d_s$-norm is at least non increasing.  Let us set
 \[
A = \lambda\beta + \mu, \quad B = \lambda\alpha + \tilde\mu.
 \]
In reason of condition \fer{co3}, both $A$ and $B$ are strictly positive.
It is immediate to recognize that
 \be\label{co4}
 A\xi^* - B\eta^* = (1-A-B)(A\xi - B\eta)
 \ee
Therefore, if the random variables $\mu$ and $\tilde\mu$, in
addition to \fer{co3} satisfy the additional constraint
 \be\label{co41}
\langle |1-A-B |^s\rangle = \langle |1-\lambda -\mu-\tilde\mu|^s
\rangle < 1
 \ee
for some positive $s$, it holds
\[
C_s= \left\langle \frac{(|\xi^*+\eta^*|^2 +
|A\xi^*-B\eta^*|^2)^{s}}{(|\xi+\eta|^2 + |A\xi-B\eta|^2)^{s}}
\right\rangle \le 1,
 \]
Hence, by passing to the new variables
 \[
\xi_1 = \xi + \eta, \quad \eta_1 = A\xi - B\eta
 \]
we can prove that, with respect to this set of variables the $d_s(f,g)$ is
not increasing for all $s >0$. This argument implies the global existence
of a unique solution to the Boltzmann equation \fer{Bl}, but it is not
conclusive with respect to the convergence of the global solution towards
a unique (in terms of the initial density) steady state. Also, it is not
clear if, along the line in which the measure steady solution is
concentrated, there is a behavior at infinity with fat tails. This is
related to the possibility that, under the constraint of conservation of
the total number of goods in a binary exchange, there is a possibility to
form a class of rich agents.

 Instead of
working directly on the Boltzmann equation \fer{Bl}, in the next
Section we will deal with the derivation of a Fokker-Planck equation
which, while keeping the main properties of the Boltzmann equation,
will be easier to treat in connection with the aforementioned
question about tails.

\section{Fokker-Planck models}\label{sec-FP}
As often happens when dealing with models of multi-agent systems,
one of the fundamental issues is to understand the emergent
properties appearing in reason of the type of interactions. In the
problem under study, this corresponds to the knowledge of the
properties of the equilibrium distribution $f_\infty(x,y)$. To this
aim, let us introduce a new set of variables, which are more adapted
to our purposes. Starting from \fer{new1},  let us define
 \be\label{nv}
v = {m_y x + m_x y} , \quad w = { m_y x - m_x y}.
 \ee
Note that $v \in \real_+$, while $w \in \real$. In addition, by
construction $|w| \le v$.  With respect to the new variables
$(v,w)$, the trade induced by the Edgeworth box reads
\begin{equations}\label{vv}
  &v^* = v + {\left[ \lambda(\alpha -\beta) + \tilde\mu -\mu \right]} w \\
    &w^* = \left( {1-\lambda} + \tilde\mu +\mu \right) w.
\end{equations}
These new variables better enlighten the  outcome of the exchange,
and the consequences we described by Fourier transform methods in
Section \ref{sec-lin}. In absence of randomness, the interaction is
dissipative in the $w$ variable, since
 \[
|w^*| = (1-\lambda)|w|.
 \]
The same property remains true if the random variables $\mu$ and
$\tilde\mu$ are such that, for a given $\lambda$
 \be\label{dis}
 \left\langle({1-\lambda} + \tilde\mu +\mu)^2\right\rangle = (1-\lambda)^2 + \left\langle(\tilde\mu +\mu)^2\right\rangle<1.
 \ee
Note that condition \fer{dis} is the analogous of condition \fer{co41},
which relates the smallness of the support of the  random variables to the
size of $\lambda$.  If \fer{dis} holds true, in view of the dissipation,
the solution to the linear Boltzmann equation \fer{Bl} will concentrate in
time on the line $m_y x = m_x y$.

As outlined at the end of Section \ref{sec-lin}, the interesting
question to answer is to describe the equilibrium profile in the
$v$-variable, to understand if the action of the interaction
\fer{vv} could produce or not Pareto tails. Since the new variables
$(v,w)$ seem more suitable to describe the problem, let us set
$g(v,w,t) = f(x,y,t)$. The condition $|w| \le v$ implies that the
new density vanishes outside the allowed set of values, so that
 \be\label{c4}
g(v,w,t) = 0, \quad {\rm if} \,\, |w| > v.
 \ee
On the set $|w| \le v$,  $g$ satisfies the linear Boltzmann equation
 \be\label{Bg}
\frac{d}{dt} \int  \varphi(v, w) g(t)dv dw = \left\langle \int  (
\varphi(v^* ,w^*)-\varphi(v ,w))g(t)dv dw \right\rangle .
 \ee
In the rest of the Section, let us assume that the random variables $\mu$
and $\tilde\mu$ are identically distributed, with zero mean and variance
$\sigma^2$. Thanks to conditions \fer{co3}, the random variables possess
in addition moments bounded of any order. Let us expand the smooth
function $\varphi(v^*,w^*)$ in Taylor series up to order two. We obtain
\begin{align}
  \nonumber
  & {\langle \varphi(v^* ,w^*)-\varphi(v ,w)) \rangle} = \\
 \nonumber
   & {\lambda}\left[ (\alpha -\beta)w\frac{\partial \varphi}{\partial v} -w \frac{\partial \varphi}{\partial w}
   \right] + \frac 12{\sigma_1^2} w^2 \frac{\partial^2 \varphi}{\partial v^2} + \frac 12{\sigma_2^2} w^2 \frac{\partial^2 \varphi}{\partial
   w^2}+ \\
   \nonumber
   & \frac 12{\lambda^2} \left[ (\alpha -\beta)^2 w^2 \frac{\partial^2\varphi}{\partial v^2} +w^2 \frac{\partial^2\varphi}{\partial
   w^2} -2(\alpha - \beta) w^2 \frac{\partial^2\varphi}{\partial v \partial
   w}
   \right] + {R(v,w)}.
\end{align}
Clearly, $R(v,w)$ denotes the remainder of the Taylor expansion.

Let us set $\tau = {\e} t$, and let us consider the situation in which $
\lambda \to {\e} \lambda$, while $\mu \to \sqrt\e \mu$ and $\tilde\mu \to
\sqrt\e \tilde\mu$. Then $h_\e(v,w,\tau) = g(v,w,t)$ satisfies
 \begin{align}
  \nonumber
  & \frac{d}{d\tau} \int  \varphi(v, w) h_\e(v,w,\tau)dv dw  = \\
 \nonumber
   & \int \left[ {\lambda}(\alpha -\beta)w\frac{\partial \varphi}{\partial v} -\lambda w \frac{\partial \varphi}{\partial w}
    + \frac 12{\sigma_1^2} w^2 \frac{\partial^2 \varphi}{\partial v^2} +  \frac 12{\sigma_2^2} w^2 \frac{\partial^2 \varphi}{\partial
   w^2}\right] h_\e(\tau) dv dw +  \\
   \nonumber
   & {\e}\frac 12{\lambda^2}\int \left[ (\alpha -\beta)^2 w^2 \frac{\partial^2\varphi}{\partial v^2} +w^2 \frac{\partial^2\varphi}{\partial
   w^2} -2(\alpha - \beta) w^2 \frac{\partial^2\varphi}{\partial v \partial
   w}
   \right]h_\e(\tau) + {R_\e(v,w)}.
\end{align}
We remark that, by construction,  the remainder $R_\e(v,w)$ depends in a
multiplicative way on higher moments of the random variables $\sqrt\e \mu$
and $\sqrt\e \tilde\mu$, so that $ R_\e(v,w)/ \e \ll 1$ for $\e\ll 1$
(cfr. the discussion in \cite{CPT, To1}, where similar computations have
been done explicitly).

 As {$\e \to 0$} $h_\e(v,w,\tau) \to h(v,w,\tau)$ satisfying
\begin{equations}\label{ww1}
  & \frac{d}{d\tau} \int  \varphi(v, w) h(v,w,\tau)dv dw  = \\
   & \int \left[ {\lambda}(\alpha -\beta)w\frac{\partial \varphi}{\partial v} -\lambda w \frac{\partial \varphi}{\partial w}
    + \frac 12{\sigma_1^2} w^2 \frac{\partial^2 \varphi}{\partial v^2} +  \frac 12{\sigma_2^2} w^2 \frac{\partial^2 \varphi}{\partial
   w^2}\right] h(\tau) dv dw.
\end{equations}
Equation \fer{ww1} is the weak form of the Fokker-Planck equation
 \be\label{FPa}
 \frac{\partial h}{\partial\tau} = \frac 12 \sigma_1^2 w^2 \frac{\partial^2 h}{\partial v^2} +
\frac 12 \sigma_2^2 \frac{\partial^2(w^2 h)}{\partial
   w^2} - {\lambda}(\alpha -\beta)w\frac{\partial h}{\partial v} + \lambda
\frac{\partial(wh)}{\partial w}.
 \ee
The coefficients in equation \fer{FPa} are given by {$\sigma_1^2 = \langle
(\mu -\tilde\mu)^2\rangle$}, and {$\sigma_2^2 = \langle (\mu
+\tilde\mu)^2\rangle$}. Clearly $\sigma_1^2= \sigma_2^2 = 2\sigma^2$ if
the errors are {uncorrelated}.

The formal derivation of the Fokker-Planck equation \fer{FPa} can be
made rigorous  repeating the analogous computations of \cite{CPT,
To1}, which refer to one-dimensional models. The meaning of this
derivation is that we allow for small changes in a single trade of
goods. Then, in order that a macroscopic change would be visible,
the system needs to wait a sufficiently long time. We call this
procedure \emph{quasi-invariant trade limit}. It is interesting to
remark that the balance {$\sigma^2/\lambda = C$} is the right one
which maintains in the limit equation both the effects of the
exchange of goods in terms of the intensity $\lambda$, as well as
the effects of the randomness (through the variance $\sigma$). As
explained in \cite{To1} for the case of opinion formation, different
balances give in the limit purely diffusive equations, of purely
drift equations.

The Fokker-Planck equation \fer{FPa} is reminiscent of all parameters
which contributed to the exchange mechanism. In particular, it contains
the values $\alpha$ and $\beta$, with $\alpha + \beta = 1$, which are the
exponent in the Cobb-Douglas utility function. It is remarkable that these
parameters appear only in terms of their difference, so that the case
$\alpha = \beta$ leads to a simplification of the equation. Also, the
values of $\lambda$ and $\sigma^2$ are linked to the differential terms of
the equation, the former to the drift, and the latter to the diffusion.

Maybe the most important characteristic of the solution to the
Fokker-Planck equation, is that it is immediate to recognize that it
develops fat tails. This property follows by evaluating moments
relative to the {$w$} variable. Let us set {$\varphi(v,w) =
|w|^{1+r}$}, with $r >0$. Then, assuming that the solution is
rapidly vanishing at infinity, one obtains
 \[
 \frac{d}{d\tau} \int |w|^{1+r}  h(v,w,\tau)dv dw  =
 \]
 \[
  \left({ \frac 12 r \sigma_2^2 -\lambda}\right)(1+r)\int |w|^{1+r} h(v,w,\tau) dv
 dw .
 \]
The sign of the right-hand side determines the large-time behavior
of the solution. If {$r \ge 2\lambda/\sigma_2^2$} the moment
$m_{1+r} = \int |w|^{1+r} h(v,w,t)dv dw $ {blows up} as time goes to
infinity. On the contrary, in agreement with the behavior of the
Fourier transformed Boltzmann equation studied in Section
\ref{sec-lin}, if {$r < 2\lambda/\sigma_2^2$} the solution
concentrates along the line $w = 0$, that is {$ m_y x = m_x y$}.
Moreover, since $h(v,w,t) = 0$ whenever $w > v$,
 \be\label{mom}
\int v^{1+r}  h(v,w,\tau)dv dw \ge \int |w|^{1+r}  h(v,w,\tau)dv dw.
 \ee
Hence, if $r \ge 2\lambda/\sigma_2^2$, also the principal moment of
order $r$ with respect to the $v$ variable blows up when time goes
to infinity, and the stationary solution, which is concentrated
along the line $w = 0$ has Pareto tails.

\section{Final remarks}
In this paper we introduced kinetic equations for the evolution of
the probability distribution of two goods among a huge population of
agents. The leading idea was to describe the trading of these goods
by means of some fundamental rules in prize theory, in particular by
using Cobb-Douglas utility functions for the binary exchange, and
the Edgeworth box for the description of the common exchange area in
which utility is increasing for both agents. Also, to take into
account the intrinsic risks of the market, we introduced randomness
in the trade, without affecting the microscopic conservations of the
trade, that is the conservation of the total number of each good in
the binary exchange.

The linear Boltzmann equation we introduced in Section \ref{sec-lin} and
its asymptotic Fokker-Planck limit studied in Section \ref{sec-FP}, showed
that, despite the microscopic constraints, sufficiently high moments of
the solution blow up with time. This is in contrast with the behavior of
the simplified one-dimensional models for wealth distribution expressed by
equation \fer{eq.boltzmann}, consequent to trades of type \fer{eq.trules},
where the microscopic conservation of wealth in the binary exchange has
been shown to imply an exponential decay of the stationary profile at
infinity \cite{DMTb, MTa}, thus excluding the possibility of Pareto tails.

The model we introduced in this paper can be generalized in many ways.
First, the binary trade \fer{tr} suggested by the Edgeworth box and
considered in Section \ref{Edge}  can be modified in many ways, which
continue to satisfy all constraints imposed by price theory. One of the
possible modifications is given by
\begin{equations}\label{tr6}
&p^* = p(1 +\mu) + {\lambda \beta (q-p) } \\
    &q^* = q(1+ \tilde\mu) + {\lambda \alpha (p-q) }.
\end{equations}
Analogously to \fer{tr}, in \fer{tr6}  $0 <\lambda \le 1$, and $\mu$ and
$\tilde \mu$ are random variables with zero mean and {finite variance},
which satisfy conditions \fer{co3}. At difference with the previous
interactions, here the randomness modifies the point in the Edgeworth box
proportionally to the percentages of agents (not to their difference). The
main consequence of this choice relies in the fact that now $p= q$ does
not imply $p^* = q^*$, but only $\langle p^*\rangle = \langle q^*\rangle$
(equality in the mean). This new way of looking at randomness reflects
heavily on the steady state of the linear Boltzmann equation \fer{Bl},
which does not concentrate along a well-defined line.

Second, the strategy of using utility functions and Hedgeworth box for the
trading is not restricted to an agent-based market in which agents
exchange two goods only. One can easily generalize the analysis of the
previous Sections to model markets in which there are $N>2$ types of goods
to be exchanged, according to \fer{trA} for each binary trade.

Also, one can improve the trading by allowing people to possess
different utility functions. This can be easily done for example by
introducing a Cobb-Douglas utility function \fer{CD} in which the
exponents $\alpha$ and $\beta$ are positive random variables,
subject to the constraint
 \[
 \alpha + \beta = 1.
 \]
This generalization will introduce a change in the Boltzmann
equation, while it will induce the same Fokker-Planck limit, with
the coefficient $\alpha- \beta$ substituted by   $\langle \alpha-
\beta\rangle$.

\medskip
{\bf Acknowledgement.} This work has been done under the activities of the
National Group of Mathematical Physics (GNFM). The support of the MIUR
project ``Variational, functional-analytic, and optimal transport methods
for dissipative evolutions and stability problems'' is kindly
acknowledged.


\medskip





\begin{thebibliography}{10}


\bibitem{angle} J.\ Angle,
   The surplus theory of social stratification and the size distribution of personal wealth
  {\em Social Forces\/} {\bf 65}(2), 293--326 (1986).

\bibitem{BMP02} { A.\ Baldassarri, U.\ Marini Bettolo Marconi and A.\ Puglisi},
\newblock Kinetic models of inelastic gases.
\newblock {\em Mat.\ Mod.\ Meth.\ Appl.\ Sci.} \textbf{12} (2002), 965--983.

\bibitem{Basu}
B. Basu, B.K. Chackabarti, S.R. Chackavarty, K. Gangopadhyay.
\emph{Econophysics \& Economics of Games, Social Choices and Quantitative
Techniques},  New Economic Windows Series, Springer, Milan, 2010.

\bibitem{BK00} { E.\ Ben-Naim and P.\ Krapivski},
\newblock Multiscaling in inelastic collisions.
\newblock {\em Phys.\ Rev.\ E} \textbf{61} (2000), R5--R8 .

\bibitem{BK02}
{ E.\ Ben-Naim and P.\ Krapivski},
\newblock Nontrivial velocity distributions in inelastic
gases.
\newblock {\em J.\ Phys.\ A} \textbf{35} (2002), L147--L152.

\bibitem{BCT06}
  M.\ Bisi, J.A.\ Carrillo and G.\ Toscani,
  Decay rates in probability metrics towards homogeneous cooling states for the inelastic Maxwell model.
  {\em J.\ Stat.\ Phys.} 124 (2-4)  (2006) 625-653.

\bibitem{Bob}
A.V.\ Bobylev,
\newblock The theory of the nonlinear spatially uniform Boltzmann equation for Maxwellian
molecules. {\em Sov.\ Sci.\ Rev.\ c} \textbf{7} (1988), 111-233.

\bibitem{Bobylev-Carrillo-Gamba}
 { A.V.\ Bobylev, J.A.\ Carrillo and I.\ Gamba},
On some properties of kinetic and hydrodynamic equations for inelastic
interactions. {\it J.\ Stat.\ Phys.} {\bf 98} (2000), 743--773; Erratum
on: {\it J.\ Stat.\ Phys.} {\bf 103}, (2001), 1137--1138.

\bibitem{BT}
C. Brugna and G. Toscani, Kinetic models for trading. A numerical
approach. (Work in progress)

\bibitem{CCT}
J.A. Carrillo, S. Cordier, G. Toscani, Over-populated tails for
conservative-in-the-mean inelastic Maxwell models, \emph{Discrete
and Continuous Dynamical Systems A}. \textbf{24} (1) (2009) 59--81

\bibitem{CT}
J.A. Carrillo and G. Toscani,  Contractive probability metrics ans
asymptotic behavior of dissipative kinetic equations.  \emph{Riv. Mat.
Univ. Parma}, (7) \textbf{6}, (2007) 75--198.

\bibitem{Cer}
C. Cercignani. \emph{The Boltzmann equation and its applications},
\newblock {Springer Series in Applied Mathematical Sciences},
  Vol.\textbf{67} Springer--Verlag 1988.

\bibitem{Cer94}
C. Cercignani, R. Illner, M. Pulvirenti,
\newblock The mathematical theory of dilute gases,
\newblock {\em Springer Series in Applied Mathematical Sciences},
  Vol.\textbf{  106} Springer--Verlag 1994.


\bibitem{CC00} A.\ Chakraborti and B.K.\ Chakrabarti,
   Statistical mechanics of money: how saving propensity affects its distributions.
  \emph{Eur.~Phys.~J.~B.\/} \textbf{17} (2000), 167--170.

\bibitem{book2} B.K.\ Chakrabarti, A.\ Chakraborti, and A.\ Chatterjee,
  {\it Econophysics and Sociophysics: Trends and Perspectives},
  Wiley VCH, Berlin, 2006.

\bibitem{review} A.\ Chatterjee and B.K.\ Chakrabarti,
   Kinetic Exchange Models for Income and Wealth Distributions.
  \emph{Eur.~Phys.~J.~B\/} {\bf 60} (2007), 135--149.

\bibitem{CCM} A.\ Chatterjee, B.K.\ Chakrabarti, and S.S.\ Manna,
   Pareto law in a kinetic model of market with random saving propensity.
  \emph{Physica A\/} {\bf 335} (2004), 155--163.

\bibitem{CCS} A.\ Chatterjee, B.K.\ Chakrabarti, and R.B.\ Stinchcombe,
   Master equation for a kinetic model of trading market and its analytic solution.
  \emph{Phys.\ Rev.\ E\/} {\bf 72} (2005), 026126.


\bibitem{book1} A.\ Chatterjee, Y.\ Sudhakar, and B.K.\ Chakrabarti,
  {\it Econophysics of Wealth Distributions}
   New Economic Windows Series, Springer, Milan, 2005.

\bibitem{CPP}
S.\ Cordier, L.\ Pareschi and C. Piatecki, Mesoscopic modelling of
financial markets, \emph{J Stat Phys} \textbf{134} 161--184 (2009).


\bibitem{CPT} S.\ Cordier, L.\ Pareschi and G.\ Toscani, On a kinetic model for a simple market economy.
  \emph{J.~Stat.~Phys.\/} {\bf 120} (2005), 253--277.


\bibitem{DY00} A.\ Dr\v{a}gulescu and V.M.\ Yakovenko,
   Statistical mechanics of money.
  \emph{Eur.~Phys.~Jour.~B\/} \textbf{17} (2000), 723--729.

\bibitem{DMT} B.\ D\"uring, D.\ Matthes and G.\ Toscani, Exponential and algebraic relaxation in kinetic models for wealth
distribution. In: {\em ``WASCOM 2007'' - Proceedings of the 14th
Conference on Waves and Stability in Continuous Media}, N.\ Manganaro et
al.\ (eds.), pp.\ 228--238, World Sci.\ Publ., Hackensack, NJ, 2008.

\bibitem{DMTb} B.\ D\"uring, D.\ Matthes and G.\ Toscani, Kinetic Equations modelling Wealth Redistribution:
A comparison of Approaches.  \emph{Phys.\ Rev.\ E} 78,   (2008) 056103.

\bibitem{DT} B.\ D\"uring and G.\ Toscani, Hydrodynamics from kinetic
  models of conservative economies.
  \emph{Physica A\/} {\bf 384} (2007), 493--506.

\bibitem{Edg}
F.Y. Edgeworth,   \emph{Mathematical Psychics: An Essay on the
Application of Mathematics to the Moral Sciences}, Kegan Paul,
London 1881.

\bibitem{Fri}
D.D. Friedman, \emph{Price Theory: An Intermediate Text}, South-Western
Publishing Co. (1990).

\bibitem{GTW95}
  E. Gabetta, G. Toscani and B. Wennberg,
  Metrics for probability distributions and the trend to equilibrium for solutions of the Boltzmann equation.
  {\em J. Statist. Phys.} \textbf{81},  901--934 (1995).

\bibitem{Gup06} K.~Gupta,
Money exchange model and a general outlook.
 \emph{Physica A} \textbf{359} (2006), 634--640.

\bibitem{hayes} B.\ Hayes,
  Follow the money.
  \emph{American Scientist\/} \textbf{90}(5) (2002), 400--405.

\bibitem{IKR} S.\ Ispolatov, P.L.\ Krapivsky and S.\ Redner.
  Wealth distributions in asset exchange models.
  {\em Eur.~Phys.~Jour.~B\/} \textbf{ 2}, 267--276 (1998).


\bibitem{KTa}
D. Kahneman and A. Tversky, Prospect theory: an analysis of decision under
risk, \emph{Econometrica}  263--292 (1979).

\bibitem{KTb}
D. Kahneman and A. Tversky, \emph{Choices, Values, and Frames}, Cambridge
University Press, Cambridge, 2000.

\bibitem{LLS}
M. Levy, H. Levy and S. Solomon, A microscopic model of the stock market:
Cycles, booms and crashes. \emph{Econ. Lett.}  \textbf{45}, 103–111
(1994).

\bibitem{LLSb}
M. Levy, H. Levy and S. Solomon, \emph{Microscopic Simulation of Financial
Markets: From Investor Behaviour to Market Phenomena}. Academic Press, San
Diego (2000).

\bibitem{LMa}
T. Lux, M. Marchesi, Volatility clustering in financial markets: a
microscopic simulation of interacting agents, \emph{International Journal
of Theoretical and Applied Finance} \textbf{3}  675--702 (2000).

\bibitem{LMb}
T. Lux, M. Marchesi, Scaling and criticality in a stochastich multi-agent
model of a financial market, \emph{Nature} \textbf{397} (11) 498--500
(1999).

\bibitem{MD}
D. Maldarella and L. Pareschi, Kinetic models for socio--economic dynamics
of speculative markets, \emph{Physica A}, \textbf{391} 715--730 (2012).

\bibitem{MTa} D.\ Matthes and G.\ Toscani, On steady distributions of
  kinetic models of conservative economies.
  \emph{J.~Stat.~Phys.} {\bf 130} (2008), 1087--1117.

\bibitem{MTb} D.~Matthes and G.\ Toscani, Analysis of a model for wealth
redistribution.
  \emph{Kinetic Rel.~Mod.} {\bf 1} (2008), 1--22.


\bibitem{NPT}
G.Naldi, L.Pareschi, G.Toscani Eds. \emph{Mathematical Modeling of
Collective Behavior in Socio-Economic and Life Sciences},  Birkhauser,
Boston (2010).

\bibitem{Pa}
L.Pareschi, G.Russo, An introduction to Monte Carlo methods for the
Boltzmann equation. \emph{ESAIM: Proceedings}, \textbf{10}, 35--75
(2001).

\bibitem{PT} L.\ Pareschi and G.\ Toscani,
   Self-similarity and power-like tails in nonconservative kinetic models.
  \emph{J.~Stat.~Phys.} {\bf 124}(2-4) (2006), 747--779.


\bibitem{slanina} F.\ Slanina,
   Inelastically scattering particles and wealth distribution in an open
   economy.
  \emph{Phys.~Rev.~E\/} {\bf 69} (2004), 046102.

\bibitem{book3} H.\ Takayasu,
  {\it Application of Econophysics},
  Springer, Tokyo, 2004.

\bibitem{book4} H.\ Takayasu,
  {\it Practical fruits of econophysics},
  Springer, Tokyo, 2005.

\bibitem{To1}
G. Toscani, Kinetic models of opinion formation, \emph{Communications in
Mathematical Sciences} \textbf{4}  481--496 (2006).

\bibitem{encyclopedia} V.M.\ Yakovenko,
  {Statistical Mechanics Approach to Econophysics}.
  In: {\it Encyclopedia of Complexity and System Science},
  R.A.\ Meyers (ed.), Springer, New York, in press.



\end{thebibliography}
\end{document}